\documentclass[aps,prl,twocolumn,superscriptaddress,showpacs,showkeys]{revtex4-1}

\usepackage{graphicx} 
\usepackage[version-1-compatibility, per-mode=fraction, output-decimal-marker={.}]{siunitx} 
\usepackage{amsmath}

\newcommand {\atrap}{\textsc{Alphatrap }}
\newcommand {\gfactor}{\textsl{\textrm{g}}-factor }
\newcommand {\boronlikeArgon}{$^{40}\rm{Ar}^{13+}$~}

\begin{document}

\title{\textsl{\textrm{g}}-factor of Boronlike Argon $^{40}\textrm{Ar}^{13+}$}

\author{I.~Arapoglou}
\email[]{ioanna.arapoglou@mpi-hd.mpg.de}
\altaffiliation{This article comprises parts of the PhD thesis work of I.~A. and H.~C., to be submitted at the Heidelberg University, Germany}
\affiliation{Max-Planck-Institut f\"ur Kernphysik, 69117 Heidelberg, Germany}

\author{A.~Egl}
\affiliation{Max-Planck-Institut f\"ur Kernphysik, 69117 Heidelberg, Germany}

\author{M.~H\"ocker}
\affiliation{Max-Planck-Institut f\"ur Kernphysik, 69117 Heidelberg, Germany}

\author{T.~Sailer}
\affiliation{Max-Planck-Institut f\"ur Kernphysik, 69117 Heidelberg, Germany}

\author{B.~Tu}
\affiliation{Max-Planck-Institut f\"ur Kernphysik, 69117 Heidelberg, Germany}

\author{A.~Weigel}
\affiliation{Max-Planck-Institut f\"ur Kernphysik, 69117 Heidelberg, Germany}

\author{R.~Wolf}
\altaffiliation[Current address: ]{ARC Centre for Engineered Quantum Systems, School of Physics, The University of Sydney, NSW 2006, Australia}
\affiliation{Max-Planck-Institut f\"ur Kernphysik, 69117 Heidelberg, Germany}

\author{H.~Cakir}
\affiliation{Max-Planck-Institut f\"ur Kernphysik, 69117 Heidelberg, Germany}

\author{V.~A.~Yerokhin}
\affiliation{Max-Planck-Institut f\"ur Kernphysik, 69117 Heidelberg, Germany}
\affiliation{Peter the Great St.~Petersburg Polytechnic University, 195251 St.~Petersburg, Russia}

\author{N.~S.~Oreshkina}
\affiliation{Max-Planck-Institut f\"ur Kernphysik, 69117 Heidelberg, Germany}

\author{V.~A.~Agababaev}
\affiliation{St.~Petersburg State University, 199034 St.~Petersburg, Russia}
\affiliation{St.~Petersburg Electrotechnical University, 197376 St.~Petersburg, Russia}

\author{A.~V.~Volotka}
\affiliation{St.~Petersburg State University, 199034 St.~Petersburg, Russia}
\affiliation{Helmholtz-Institut Jena, 07743 Jena, Germany}
\affiliation{GSI Helmholtzzentrum f\"ur Schwerionenforschung GmbH, 64291 Darmstadt, Germany}

\author{D.~V.~Zinenko}
\affiliation{St.~Petersburg State University, 199034 St.~Petersburg, Russia}

\author{D.~A.~Glazov}
\affiliation{St.~Petersburg State University, 199034 St.~Petersburg, Russia}

\author{Z.~Harman}
\affiliation{Max-Planck-Institut f\"ur Kernphysik, 69117 Heidelberg, Germany}

\author{C.~H.~Keitel}
\affiliation{Max-Planck-Institut f\"ur Kernphysik, 69117 Heidelberg, Germany}

\author{S.~Sturm}
\affiliation{Max-Planck-Institut f\"ur Kernphysik, 69117 Heidelberg, Germany}

\author{K.~Blaum}
\affiliation{Max-Planck-Institut f\"ur Kernphysik, 69117 Heidelberg, Germany}
\date{\today}

\begin{abstract}
We have measured the ground-state \textsl{\textrm{g}}-factor of boronlike argon $^{40}\textrm{Ar}^{13+}$ with a fractional uncertainty of \SI{1.4e-9}{} with a single ion in the newly developed \textsc{Alphatrap} double Penning-trap setup. The here obtained value of $\textsl{\textrm{g}}=0.663\,648\,455\,32(93)$ is in agreement with our theoretical prediction of 0.663\,648\,12(58). The latter is obtained accounting for quantum electrodynamics, electron correlation, and nuclear effects within the state-of-the-art theoretical methods. Our experimental result distinguishes between existing predictions that are in disagreement, and lays the foundations for an independent determination of the fine-structure constant.

\end{abstract}

\maketitle

The \gfactor of the bound electron permits high-precision tests of quantum electrodynamics (QED) in strong Coulomb fields. With an appropriate choice of the element and charge state, different effects can be individually addressed. The currently most stringent test of QED in strong fields has been performed with hydrogenlike silicon~\cite{Sturm2011b,Sturm2013}. The QED theory thus confirmed has been used subsequently for the determination of the electron mass~\cite{Sturm2014}, determining the current CODATA value~\cite{Codata2014}. Measuring the isotope shift of the \gfactor of lithiumlike calcium~\cite{Kohler2016} gave access to the relativistic nuclear recoil effect, scrutinising QED beyond the external-field approximation~\cite{Shabaev2017}. Finally, relativistic many-electron correlations were investigated using lithiumlike silicon~\cite{Wagner2013}.

The experimental determination of the \gfactor of a boronlike ion allows, for the first time, for precision tests of QED involving a bound electron possessing orbital angular momentum, and for more stringent tests of many-electron correlations. Furthermore, such ions can also be used in the future for an independent determination of the fine-structure constant $\alpha$~\cite{Shabaev2006,Volotka2014,Yerokhin2016}, competitive in precision with the presently best literature value~\cite{Parker2018}.

\begin{figure*}[ht]
    \includegraphics[keepaspectratio=true, width=\textwidth]{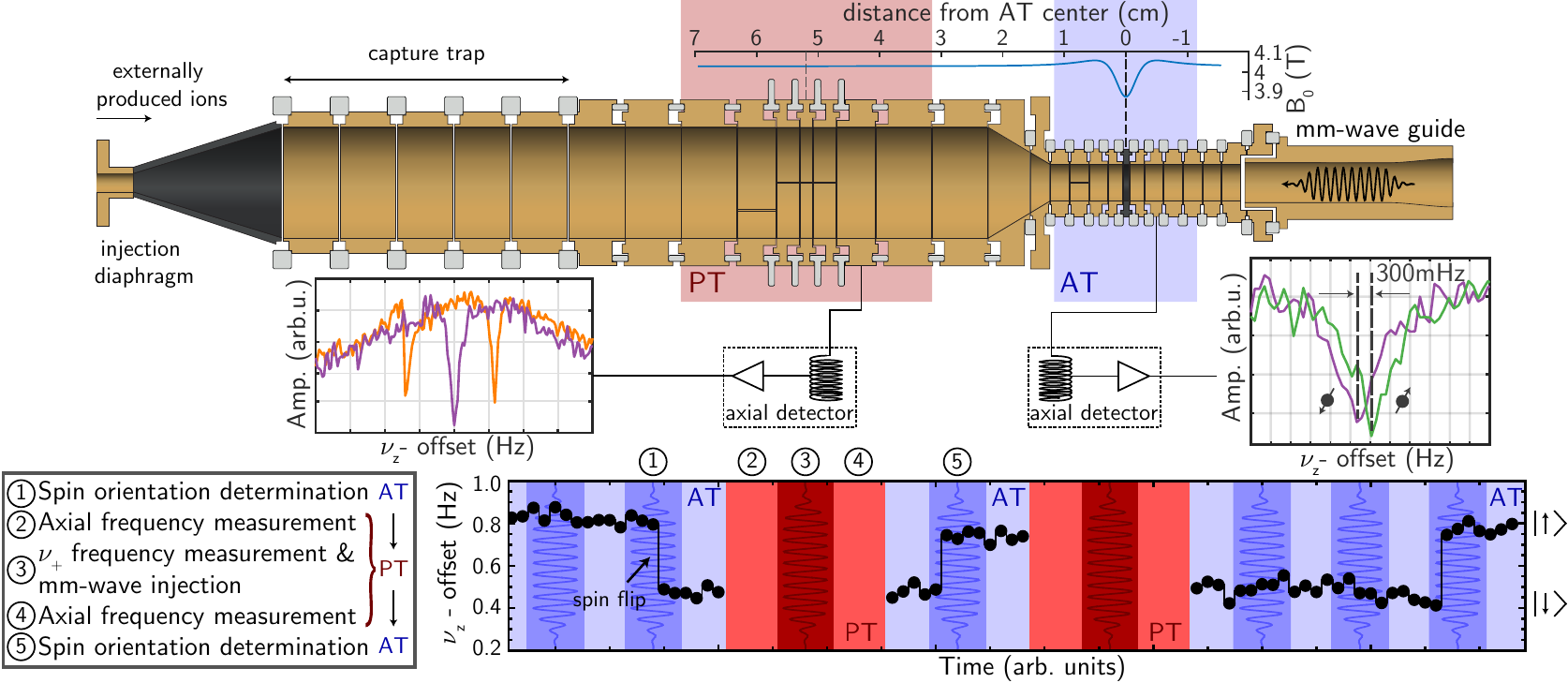}
    \caption{ The \atrap double-trap system consists of the precision trap (PT) used for high-precision spectroscopy and the analysis trap (AT) for spin-state determination. At the end of the trap tower the millimeter-wave guide is attached. After the externally created ion bunch is decelerated by a pulsed drift tube and dynamically captured by rapid switching of the potential applied on the electrodes of the capture trap, it is transported to the double Penning-trap system shown here, specifically, to the PT. There, the ion cloud is reduced to a single $^{40}\rm{Ar}^{13+}$. The measurement cycle is described in the lower part of the figure.}
    \label{fig:trap}
\end{figure*}

In this Letter we present the first result of the \textsc{Alphatrap} experiment, a Penning-trap setup for high-precision determination of \textsl{\textrm{g}}-factors. We have measured the \gfactor of the $1s^22s^22p_{1/2}$ ground-state $^{40}\rm{Ar}^{13+}$, which has been inaccessible to the previous Penning-trap experiment~\cite{Kohler2015}, and compared it with theoretical predictions. The precision of this measurement allows for testing all of the presently accessible contributions to the theoretical value. It also has great potential for future tests of higher-order contributions, which have not been calculated yet. Consequently, this measurement paves the way to perform bound-state QED tests with \atrap in even stronger fields, ultimately with highly charged lead ions, and is an important contribution towards $\alpha$ determination with heavy highly charged ions (HCI)~\cite{Shabaev2006,Volotka2014}. Additionally, we present a theoretical calculation with improved accuracy of the \textsl{\textrm{g}}-factor. The uncertainty of the one-loop QED contribution has been reduced by a factor of three. The electron-correlation contribution has been recalculated using two independent methods: large-scale configuration-interaction method in the Dirac-Fock-Sturm basis (CI-DFS) and recursive perturbation theory. The current relative uncertainty of the theoretical \gfactor is \SI{9e-7}{} and is dominated by the uncertainty of the many-electron QED and nuclear recoil terms. Still, the theoretical uncertainty is almost three orders of magnitude larger than the experimental one, making further improvement of the theory highly anticipated.

\textsc{Alphatrap}, which is the follow-up experiment to the Mainz \gfactor experiment on HCI~\cite{Sturm2014,Sturm2013,Kohler2016}, allows the injection of externally produced ions up to hydrogenlike lead. A detailed description of \atrap can be found in Ref.~\cite{alphatrapReview}. A double Penning-trap system is inserted into the bore of a \SI{4.02}{\tesla} superconducting magnet. A liquid helium tank cools the trap as well as the detection electronics to \SI{4.2}{\kelvin}. Owing to the integration of a cryogenically operable valve, the vacuum inside the trap is better than \SI{e-16}{\milli\bar} despite the external coupling, and ensures the absence of disturbances in the ion motion due to collisions and allows for virtually unlimited ion storage time. The trap is connected via a beamline to several ion sources including a Heidelberg Compact Electron Beam Ion Trap~\cite{Micke2018} that gives access to medium-\textit{Z} HCI (with $Z$ being the atomic number) and the cryogenic high-energy Heidelberg electron beam ion trap~\cite{Crespo2004}, which enables access to the high-\textit{Z} regime for novel experiments.

 Determining the \gfactor requires measuring the Larmor frequency $\nu_{\rm{L}}=\textsl{\textrm{g}}B_0 e/ (4\pi m_{\rm{e}})$, where $e$ and $m_{\rm{e}}$ denote the electron's charge and mass, respectively, in a well-known magnetic field $B_0$. The latter is deduced via the measurement of the ion's free-cyclotron frequency $\nu_{\rm{c}}=qB_0/(2\pi M)$, where $q$ and $M$ are the ion's charge and mass, respectively. While $\nu_{\rm{c}}$ is being determined, the ion is simultaneously irradiated with millimeter waves at frequencies $\nu_{\rm MW}$ close to the Larmor frequency $\nu_{\rm{L}}$.  The Larmor frequency is extracted from measuring the spin-flip probability for different excitation frequencies $\nu_{\rm{MW}}$. The \gfactor is obtained from

 \begin{equation}
 \textsl{\textrm{g}} = 2\frac{\nu_{\rm{L}}}{\nu_{\rm{c}}} \frac{q}{e} \frac{m_{\rm{e}}}{M} = 2\Gamma_0 \frac{q}{e} \frac{m_{\rm{e}}}{M},
 \label{eq:gfactor}
 \end{equation}
 
\noindent where $\Gamma_0$ denotes the frequency ratio $\nu_{\rm{L}}/\nu_{\rm{c}}$.

The single ion's motion in a Penning trap is a superposition of three independent harmonic oscillation modes with the modified cyclotron frequency $\nu_+\approx\SI{20}{\mega\hertz}$, the axial frequency $\nu_{\rm{z}}\approx\SI{650}{\kilo\hertz}$ and the magnetron frequency $\nu_-\approx\SI{10}{\kilo\hertz}$ in our setup. The free-cyclotron frequency of the ion is determined by means of the Brown-Gabrielse invariance theorem $\nu_{\rm{c}}^2~=~\nu_+^2~+~\nu_{\rm{z}}^2~+~\nu_-^2$~\cite{Brown1982}, where frequency shifts due to possible tilts and elliptic deformations of the trapping potential are canceled. 

These frequencies are detected non-destructively by measuring the ion-induced image current on axially separated electrodes. The ion's oscillation in the axial direction is brought into resonance with a cryogenic superconducting tank circuit with a quality factor of $Q = 38500$.  The voltage drop across the impedance is Fourier transformed and the ion's frequency appears as a minimum in the noise spectrum of the detection circuit, the so-called ``dip'' signal as shown in the insets of Fig.~\ref{fig:trap}. After resistive cooling, the ion eventually reaches thermal equilibrium with the tank's effective mode temperature. In our setup this temperature amounts to about \SI{6}{\kelvin}, slightly above the ambient temperature of \SI{4.2}{\kelvin}. In addition to the axial frequency, the two radial frequencies are detected on the axial detector by coupling them to the axial mode via a radio-frequency sideband drive at frequencies $\nu_+ - \nu_{\rm{z}}$ and $\nu_{\rm{z}} + \nu_-$~\cite{Cornell1990,Sturm2010}. This forces the coupled modes into a Rabi oscillation, which splits the dip in the noise spectrum into two dips, the so-called ``double-dip''. This way, the determination of the modified cyclotron and the magnetron frequency becomes possible.

\begin{figure}[t]
    \includegraphics[keepaspectratio=true, width=\columnwidth]{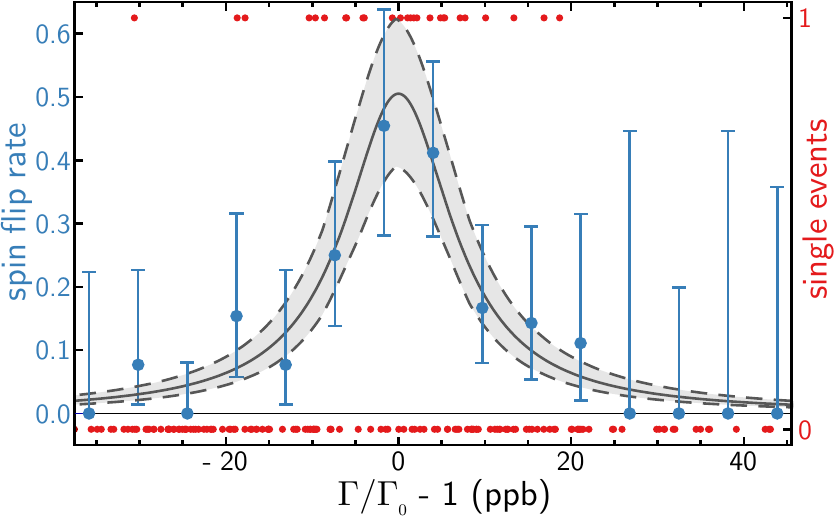}
    \caption{$\Gamma$ resonance (Res.~B in Table I), depicting the spin-flip probability as a function of the frequency ratio $\Gamma = \nu_{\rm MW}/\nu_{\rm{c}}$. The data is fitted  with a Lorentzian (solid line) using the maximum likelihood method. The dashed lines indicate the 1-$\sigma$ confidence interval of the fit. The blue points represent the binned data with binomial error bars and are included in the plot only as a guide for the eye. The red dots represent the single spin-flip events with 1 being a successful spin flip and 0 being an unsuccessful one.}
    \label{fig:gammaResonance}
\end{figure}

In order to additionally measure the Larmor frequency, a typical experimental cycle is as follows (see also Fig.~\ref{fig:trap}): The ion is adiabatically transported to the analysis trap, where the ion's spin state is determined by means of the continuous Stern-Gerlach effect~\cite{Werth2002}. The strength of the magnetic bottle introduced by a ferromagnetic ring in the analysis trap has been measured to be $B_2 = \SI{44.35(84)}{\kilo\tesla\slash\square\meter}$. This quadratic inhomogeneity creates an additional axial force $\vec{F} = 2\mu_{\rm{z}} B_2 \hat{z}$ which depends on the magnetic moment orientation. Within this configuration a spin-state change is observed as an axial frequency jump, which for \boronlikeArgon corresponds to $\Delta\nu_{\rm{z}} = \SI{312(6)}{\milli\hertz}$ out of \SI{335}{\kilo\hertz} axial frequency. Therefore, for unambiguous spin-flip detection the ion trapping voltage needs to be stable at a level of $\delta U/U \leq \SI{4.5e-7}{}$. After probing the spin state, the ion is adiabatically transported to the precision trap. There, the ion is irradiated with millimeter waves at a frequency near the Larmor frequency $\nu_{\rm{L}} \approx \SI{37}{\giga\hertz}$. Simultaneously, the ion's free-cyclotron frequency $\nu_{\rm{c}}$ is measured, allowing the determination of the ratio  $\Gamma=\nu_{\rm MW}/\nu_{\rm{c}}$. For this, all three eigenfrequencies of the ion are measured within the highly homogeneous magnetic field of the precision trap which avoids adverse line-broadening effects of the magnetic bottle. Finally, the ion is transported back to the analysis trap for determining whether a spin flip occurred during the millimeter-wave irradiation in the precision trap. Repeating this measurement cycle several times results in a resonance of the spin-flip probability as a function of the corresponding $\Gamma$ ratio as shown in Fig.~\ref{fig:gammaResonance}.

Due to the comparably low precision of the theoretical prediction ($\sim$\SI{}{ppm}) compared to the typical line-width of the experiment (\SI{<10}{ppb}), we have used an adiabatic rapid passage~\cite{Camparo1984} measurement scheme for the initial resonance search. To this end, the magnetic field was swept using a set of Helmholtz coils that was installed outside the superconducting magnet. In combination with the background-free spin-state detection, this method allows an efficient search in a comparably large frequency range.

After the resonance was found, it has been recorded twice (Res.~A and Res.~B in Table~\ref{tab:syst}), with a slightly improved measurement sequence used for Res.~B. Using a maximum likelihood estimation, the centre of each of the resonances is determined by fitting a Lorentzian lineshape to the data. The centers of the two resonances are weighted by their individual statistical uncertainty and the uncertainty of the axial potential drift corrections (see Table~\ref{tab:syst}). This yields $\Gamma_0' = 1859.082\,876\,9(23)$ with a relative statistical uncertainty of \SI{1.26e-9}{}. This value needs to be corrected for systematic shifts. The dominant effect during this measurement campaign was a drift of the axial frequency in the precision trap during the $\Gamma$-ratio determination arising from the slow thermalisation of the power supply when the trapping voltages are set. The applied voltages ($\approx\SI{-75}{\volt}$) are almost by an order of magnitude larger than previous experiments, making this effect non-negligible. This effect, which is caused by voltage changes during ion transport, has been determined with a dedicated measurement to yield the values of Table~\ref{tab:syst}. It was significantly reduced for Res.~B by a more suitable choice of transport voltages and will be mitigated with a dedicated transport power supply for the next measurement campaign. Moreover, due to the highly optimised design of our precision trap, which includes a larger diameter of \SI{18}{\milli\meter}, the effect of electric and magnetic field imperfections are negligible. The residual inhomogeneities of the magnetic field ($B_{\rm z}=B_0+B_1z+B_2z^2+...$)~\cite{Ketter2014} along the axis of the precision trap amount to  $B_1 = \SI{2.566(29)}{\milli\tesla\slash\meter}$ and $B_2 = \SI{0.0643(32)}{\tesla\slash\meter\square}$. Furthermore, the image charge shift, which was a dominant systematic uncertainty in past experiments, is now calculated using finite element method to be $\delta\nu_{\rm{c}}/\nu_{\rm{c}}=\SI{5.03(25)e-11}{}$ for \boronlikeArgon in \atrap and is virtually negligible at the current precision level. The \SI{5}{\percent} uncertainty corresponds to a conservative estimation, which can be reduced with a more rigorous calculation if necessary. We also estimate a conservative systematic uncertainty due to the frequency pulling effect of the detuned cyclotron tank circuit. The experimental frequency ratio, corrected for all shifts given in Table~\ref{tab:syst}, is $\Gamma_0=1859.082\,876\,8(23)$. 

\begin{table}[h]
\caption{Relative systematic shifts ($(\Gamma_0-\Gamma_0')/\Gamma_0'$) and their uncertainties for each of the obtained $\Gamma$ resonances. The corresponding \gfactor at \SI{4.02}{\tesla} and the one corrected for the cubic Zeeman shift are listed below. The numbers in parentheses correspond to the statistical and systematic uncertainties, and the uncertainty due to the external constants, respectively.}
 \begin{ruledtabular}
\begin{tabular}{lccc}
\textbf{Effect}   						    	&\textbf{Res.~A}\textbf{(ppt)}		& \textbf{Res.~B}\textbf{(ppt)}\\
\hline
Drift of axial potential 					& 0(870) 			& 0(360)		\\
Image charge							&\multicolumn{2}{l}{\phantom{0000000.}$-$50.3(2.5)}\\
Relativistic mass increase					&\multicolumn{2}{l}{\phantom{00000000.}$-$0.43(6)}\\
Lineshape of dip fit						&\multicolumn{2}{l}{\phantom{0000000000}0(270)}\\
Frequency pulling 						&\multicolumn{2}{l}{\phantom{0000000000}0(50)}\\
$\nu_-$ measurement					&\multicolumn{2}{l}{\phantom{0000000000}0.0(3.4)}\\
Elevated $E_+$ during $\nu_z$ meas.		&\multicolumn{2}{l}{\phantom{0000000000}0.00(82)}\\
Electric field anharmonicity   				&\multicolumn{2}{l}{\phantom{0000000000}0.00(60)}\\
Lineshape of $\Gamma$ resonance			& \multicolumn{2}{l}{\phantom{0000000000}0.00(2)}\\
Magnetic field inhomogeneity  			 	&\multicolumn{2}{l}{\phantom{0000000000}$\ll$\SI{e-2}{}}\\
\hline
$\textsl{\textrm{g}}_{\text{exp}}$ at $B_0=\SI{4.02}{T}$	& \multicolumn{2}{c}{0.663\,648\,456\,29(83)(42)(5)}\\
$\textsl{\textrm{g}}_{\text{exp}}$ at $B_0 =\SI{0}{T}$	& \multicolumn{2}{l}{\phantom{--}0.663\,648\,455\,32(83)(42)(5)}

\end{tabular}
\label{tab:syst}
\end{ruledtabular}
\end{table}

The \gfactor is determined using eq.~(\ref{eq:gfactor}) with the electron mass $m_{\rm{e}}=\SI{5.485\,799\,090\,70(16)e-4}{\amu}$ as given by CODATA~\cite{Codata2014} and $M(^{40}\rm{Ar}^{13+})=\SI{39.955\,255\,154\,5(26)}{\amu}$. The latter is deduced after correcting the atomic mass $M(^{40}\rm{Ar})=\SI{39.962\,383\,123\,8(24)}{\amu}$~\cite{Codata2017} for the mass and binding energies of the missing electrons~\cite{NIST_ASD}. Our experimental result for the \gfactor is $\textsl{\textrm{g}}_{\text{exp}}~=~0.663\,648\,456\,29(83)(42)(5)$, where the number in the first bracket represents the statistical uncertainty, the second the systematic uncertainty and the third one accounts for the uncertainty of the electron and the argon atomic masses. 

In $^{40}\rm{Ar}^{13+}$, mixing of the closely spaced $2p_{1/2}$ and $2p_{3/2}$ levels leads to nonlinear contributions to the Zeeman splitting. However, the quadratic Zeeman shift is identical for both $m=\pm{1}/{2}$ sublevels, therefore, its contribution to the Zeeman splitting vanishes for the ground state. The lowest non-zero nonlinear term is the cubic one, $\sim B^3$. Its contribution to the \gfactor has been evaluated in Refs.~\cite{Lindenfels2013,Varentsova2018} and amounts to $6.0\times10^{-11}~(B/{\rm T})^2$. For the magnetic field of $B_0 \approx \SI{4.02}{\tesla}$ of \atrap this results in an absolute shift of \SI{9.7e-10}{}~\footnote{Based on~\cite{Lindenfels2013,Varentsova2018} and given our experimental \gfactor uncertainty, the uncertainty of these figures is negligible}. Taking into account the latter, we finally obtain for $B_0=\SI{0}{\tesla}$: 
\begin{equation}
\textsl{\textrm{g}}~=~0.663\,648\,455\,32(83)(42)(5). 
\label{eq:gfactor_final}
\end{equation}

For the theoretical \gfactor evaluation, a treatment based on the Dirac equation is necessary, including the negative-energy states and the Breit contributions to the electron-electron interaction. We take into account electron-correlation effects by means of the CI-DFS approach~\cite{Glazov2004} as in Refs.~\cite{Glazov2013,Shchepetnov2015}, confirming the results therein. This contribution has also been confirmed recently within the coupled-cluster method~\cite{Maison2019} and within second-order perturbation theory in $1/Z$~\cite{agababaev:18:jpcs}. Here, we evaluate it to higher numerical precision using the combination of the CI-DFS approach and recursive perturbation theory (P. Th.) to third and higher orders~\cite{Glazov:19:arxiv}. The contribution of the negative-energy part of the Dirac spectrum, which was found to be relevant in the case of lithiumlike ions in Ref.~\cite{Wagner2013}, is also significant here. In addition, the one-photon exchange correction is calculated in a QED framework~\cite{Shabaev2002} with a basis set constructed from B-splines within the dual kinetic-balance (DKB) approach~\cite{Shabaev2004Dual} as implemented in~\cite{Sikora2018}. In Table~\ref{tab:theory}, the results for the electron-electron interaction are presented along with subsequent terms.

\begin{table}[h]
\caption{\label{tab:theory}
Theoretical contributions to the \gfactor of ${}^{40}{\rm Ar}^{13+}$. The parenthesised numbers indicate the uncertainty of the last digit(s).
All digits are significant if no uncertainty is given.}
\label{tab:table1}
\centering
\begin{ruledtabular}
\begin{tabular}{llll}
Contribution                                  & Value                                     & Ref. \\
\hline

Dirac value                             &          \phantom{$-$}0.663\,775\,45                  &  \\

Finite nuclear size                           &  \phantom{$-0.00$}$<10^{-10}$             &      \\
Electron correlation:                         &                                           &      \\
\phantom{aa}one-photon exchange, $(1/Z)^1$    &                    \phantom{$-$}0.000\,657\,53                     &      \\
\phantom{aa}$(1/Z)^{2+}$, CI-DFS       & $-$0.000\,007\,5(4)                 & \cite{Shchepetnov2015} \\
\phantom{aa}$(1/Z)^{2+}$, P.~Th. \& CI-DFS  & $-$0.000\,007\,57(20)*		&  \\

Nuclear recoil                                &         $-$0.000\,009\,09(19)               & \cite{Glazov2018,Shchepetnov2015}  \\
One-loop QED:                                 &                                           &      \\
\phantom{aa}self-energy, $(1/Z)^{0}$          &         $-$0.000\,768\,372\,3(3)             &      \\
\phantom{aaSelf-energy,} $(1/Z)^{1+}$         &         $-$0.000\,000\,98(15)               &      \\
\phantom{aaSelf-energy,}  $(1/Z)^{1+}$    &         $-$0.000\,001\,04(19)*              &      \\
\phantom{aa}vacuum polarization               &                                           &      \\
\phantom{aaaa}electric loop, $(1/Z)^{0}$      &  \phantom{$0.$}$-4.187\times 10^{-10}$    &      \\
\phantom{aaaaelectric loop,} $(1/Z)^{1}$      &  \phantom{$-0.$}$6.526(2)\times 10^{-9}$  &      \\
\phantom{aaaa}magnetic loop, $(1/Z)^{0}$      &  \phantom{$-0.$}$4.131\times 10^{-10}$    &      \\
\phantom{aaaamagnetic loop,} $(1/Z)^{1}$      &  \phantom{$0.$}$-1.341\times 10^{-10}$    &      \\
Two-loop QED, $(Z\alpha)^0$                   & \phantom{$-$}0.000\,001\,18(6)               & \cite{Grotch1973}      \\
\hline
Total theory                                        & {\phantom{$-$}0.663\,648\,2(5)}             &    TW  \\
                                              & {\phantom{$-$}0.663\,648\,08(58)*}         &    TW  \\
Experiment			& {\phantom{$-$}0.663\,648\,455\,32(93)}  &TW \\
\end{tabular}
\end{ruledtabular}
\end{table}

The leading QED effect is due to the self-energy (SE) vertex and wave-function corrections of the $2p_{1/2}$ valence electron. In the leading $(Z\alpha)^0$ approximation, it is equal to $-\alpha/(3\pi)$~\cite{Brodsky1967}. The one-electron SE binding correction was calculated to all orders in $Z\alpha$ in Refs.~\cite{Yerokhin2008,Yerokhin2010} for $Z\le 12$. In Ref.~\cite{Glazov2013} it was calculated for $Z = 18$ with an effective screening potential. In the present work, we calculate it with an improved uncertainty using two independent methods, with the screening effect on the SE of the valence electron accounted for by means of an effective potential induced by the core electrons. Within the first method, the SE correction in a local screening potential is calculated by generalizing the numerical approach developed in Ref.~\cite{Yerokhin2013}. Computations are performed with the localized Dirac-Fock potential, the Kohn-Sham potential, and the core-Hartree potential (see, e.g., Ref.~\cite{Yerokhin2012} for details), with the result of $-769.35\,(15) \times 10^{-6}$. The uncertainty estimates the dependence on the choice of the potential and the error due to the truncation of the partial-wave expansion. Within the second method, it is calculated on the basis of the DKB finite basis set with the core-Hartree, Kohn-Sham, Dirac-Hartree, and Dirac-Slater potentials following Refs.~\cite{glazov:06:pla,volotka:06:epjd}, with the result of $-769.41\,(19) \times 10^{-6}$ (marked by * in Table~\ref{tab:theory}), in full agreement with the first method. The contribution of the two-electron SE diagrams not approximated by the above screening potential method is unknown. It can be estimated based on the corresponding calculations for lithiumlike ions~\cite{Volotka2014b}, leading to an uncertainty of $0.51\times 10^{-6}$, included in the uncertainty of the final theoretical result.

One- and many-electron vacuum polarization (VP) corrections are also evaluated. We calculate these diagrams employing a B-spline basis set. One- and two-electron magnetic-loop terms are evaluated following \cite{Lee2005,Lee2007}. In case of the \boronlikeArgon ion, these terms do not contribute at the current level of theoretical uncertainty, however, they will be important for near-future experiments with high-$Z$ systems, especially for the projected determination of $\alpha$ with very heavy ions~\cite{Shabaev2006}. Specifically, the VP terms treated here contribute as much as $-4.06\times 10^{-6}$ in ${}^{208}{\rm Pb}^{77+}$. Additionally, two-loop QED effects known only to zeroth order in $(Z\alpha)$~\cite{Grotch1973,Brodsky1967} contribute at the $10^{-6}$ level in $^{40}\rm{Ar}^{13+}$. The nuclear recoil effect in middle-$Z$ boronlike ions was evaluated to zeroth and first orders in $1/Z$ in Refs.~\cite{Shchepetnov2015,Glazov2018}. A combination of the two total theoretical results in Table~\ref{tab:theory} yields 0.663\,648\,12(58).

\begin{figure}[h!]
    \includegraphics[keepaspectratio=true, width=\columnwidth]{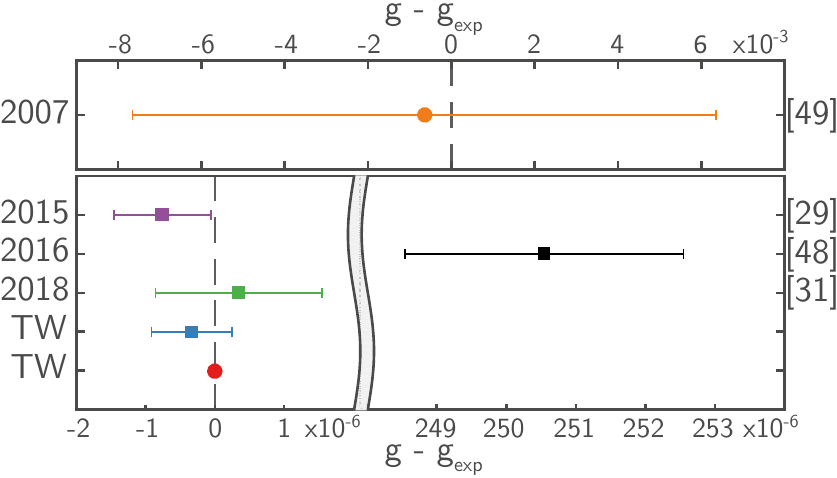}
    \caption{Comparison of the experimental (circles) and theoretical (squares) \textsl{\textrm{g}}-factors obtained in this work (TW) with previously calculated values\protect\nocite{Marques2016} as well as the previous experimental result~\cite{Soria2007} with \SI{1e-2}{} relative uncertainty (note axis label above). The fractional uncertainty of \SI{1.4e-9}{} of this work's experimental \gfactor is not visible in this plot.}
    \label{fig:gexp_gtheo_comparison}
\end{figure}

Comparing the experimental and the theoretical \gfactor values demonstrates an excellent agreement at a $10^{-7}$ level. Further improvement of the theory towards the experimental precision level will constitute a more precise test of the relativistic and QED many-electron effects. The current experimental result compared to previous calculations~\cite{Shchepetnov2015,agababaev:18:jpcs,Marques2016} as well as the improved value obtained within this work can be seen in Fig.~\ref{fig:gexp_gtheo_comparison}. It should be noted that another theoretical prediction has been published without error bars~\cite{Verdebout2014} giving a value of \textsl{\textrm{g}} = 0.663\,728. 

In summary, the first high-precision measurement of a boronlike ion's \textsl{\textrm{g}}-factor, namely that of $^{40}\rm{Ar}^{13+}$, with a fractional uncertainty of \SI{1.4e-9}{} has been presented. This level of precision is not only sufficient to test the presently available theoretical results for the electron-correlation, QED, and nuclear-recoil effects, but also to test the foreseen developments in this field, including higher-order (two-loop and many-electron) QED contributions. Theoretical calculations improved the one-loop QED contributions by a factor of 3, resulting in a total relative uncertainty of \SI{9e-7}{}. The agreement between theory and experiment represents one of the most accurate tests of many-electron QED contributions in strong fields, and paves the way towards an independent determination of the fine-structure constant.

\begin{acknowledgments}
We acknowledge financial support from the Max Planck Society. This work is part of and supported by the German Research Foundation (DFG) Collaborative Research Centre ``SFB 1225 (ISOQUANT)''. V.A.Y. acknowledges support by the Ministry of Education and Science of the Russian Federation Grant No.~3.5397.2017/6.7. The work of V.A.A., D.A.G., A.V.V., and D.V.Z. was supported in part by RFBR (Grants No.~16-02-00334 and 19-02-00974), by DFG (Grant No.~VO 1707/1-3), and by SPbSU-DFG (Grants No.~11.65.41.2017 and No.~STO 346/5-1).
\end{acknowledgments}

\bibliography{references.bib}

  \end{document}